\documentclass[galaxies,article,accept,oneauthor,LaTeX and dvi2pdf]{Definitions/mdpi} 
\firstpage{1} 
\pubvolume{xx}
\issuenum{1}
\articlenumber{5}
\pubyear{2020}
\copyrightyear{2020}
\history{Received: 30 June 2020; Accepted: 24 July 2020; Published: date}

\usepackage{subcaption}
\usepackage[countmax]{subfloat}
\Title{\textbf{\textit{NuSTAR}} View of TeV Blazar Mrk 501}

\Author{Ashwani Pandey \orcidA{}}

\AuthorNames{Ashwani Pandey}

\address[1]{Aryabhatta Research Institute of Observational Sciences (ARIES), Manora Peak, Nainital 263001, India;  ashwanitapan@gmail.com}

\abstract{We report the results of flux and spectral variability studies of all seven {\it Nuclear Spectroscopic Telescope Array (NuSTAR)} observations of TeV $\gamma-$ray emitting blazar Markarian (or Mrk) 501. We~found strong evidence of intraday variability in 3--79 keV X-ray light curves (LCs) of Mrk 501 during four out of these seven observations. We examined spectral variability using a model-independent hardness-ratio analysis and found a general ``harder-when-brighter'' behaviour in two observations. We also investigated the nature of 3--79 keV X-ray spectra of TeV blazar Mrk 501 and found that five out of seven spectra are well described by the curved log-parabola models with photon indices (at 10 keV) $\alpha \sim$ 2.12--2.32 and a curvature $\beta \sim$ 0.15--0.28. The two other spectra are somewhat better represented by simple power-law models with photon indices 2.70 and 2.75. We~briefly discuss available physical models to explain our results.}

\begin{document}
\section{Introduction} \label{sec:intro}
Active galactic nuclei (AGN) are highly energetic astrophysical objects that are powered by actively accreting supermassive black holes (SMBH; M$_{BH}$ $\geq 10^6$ M$_{\rm sun}$) at their centres, and~produce unique observational features over the entire electromagnetic (EM) spectrum \citep{1984ARA&A..22..471R}. About 15\% of AGN consist of two well-collimated jets of ultrarelativistic particles extending to kilo-parsec (kpc) or sometimes up to mega-parsec (Mpc) distances and are known as jetted-AGN \citep{2017A&ARv..25....2P}. The~spectra of these jetted-AGN strongly depend on the viewing angles of their jets with respect to the observer's line of sight and~this difference forms the base for their classification. Blazars are the jetted-AGN aligned at very small ($\leq$ 15-20$^{\circ}$) viewing angles \citep{1995PASP..107..803U, 2018ApJ...866..137L, 2019SCPMA..62l9811X}. Their spectra are fully dominated by non-thermal radiation from the relativistic jets. Blazars are usually subdivided into flat-spectrum radio quasars (FSRQs) and BL Lac objects (BLLs) \citep[e.g.][]{1991ApJS...76..813S, 1996MNRAS.281..425M}. The~division between these two subclasses could be due to their different accretion regimes with FSRQs having higher accretion rates above 10$^{-2}$ of the Eddington rate~\citep{2009MNRAS.396L.105G}. 

Blazars emit detectable radiations at all EM wavelengths and their broadband spectral energy distributions (SEDs) have characteristic double-humped structures. The~first or low energy SED hump, which peaks in Infra-red (IR) to X-rays, is dominated by the synchrotron radiation from the jet. The~second or high energy SED hump, peaking at $\gamma-$rays (GeV to TeV energies), is produced either by the inverse-Compton (IC) radiation (leptonic model;~\cite{2007Ap&SS.307...69B}) or via radiation arising from hadronic processes (hadronic model;~\cite{2003APh....18..593M}). Based on the peak frequency of the first SED hump or the synchrotron peak frequency $\nu_{\rm syn}$ \cite{2010ApJ...716...30A} subdivided {blazars into LSP (or low synchrotron peaked blazars; $\nu_{\rm syn} \leq$ 10$^{14}$ Hz), ISP (or intermediate synchrotron peaked blazars; 10$^{14} < \nu_{\rm syn} <$ 10$^{15}$ Hz), and~HSP (or high synchrotron peaked blazars; $\nu_{\rm syn} \geq$ 10$^{15}$ Hz). A slightly modified range of $\nu_{\rm syn}$ for these subclasses of blazars was proposed by \cite{2016ApJS..226...20F} as: LSP having log $\nu_{\rm syn} \rm(Hz) \leq$ 14.0, ISP with 14.0 < log $\nu_{\rm syn} \rm (Hz) \leq $ 15.3, and HSP having log $\nu_{\rm syn} \rm(Hz) > $ 15.3.}

Markarian (or Mrk)  501 ($\alpha_{\rm 2000}$ = 16h53m52s; $\delta_{\rm 2000} =	+39^{\circ}45^{\prime}37^{\prime\prime}$) is one of the nearest ($z= 0.034$; \citep{1975ApJ...198..261U}) bright X-rays emitting high-frequency peaked BL Lac object (HBL\footnote{HBL belongs to HSP subclass.}). 
 It was first detected at very high energy (VHE) $\gamma-$ ray (above 300 GeV) in 1995 by the Whipple telescope \citep{1996ApJ...456L..83Q} that made it the second extragalactic object, after~Mrk 421 \citep{1992Natur.358..477P}, detected at TeV energies. Mrk 501 has been extensively observed in several multiwavelength campaigns to understand its exact nature which is yet not fully understood (e.g., \citep{2015ApJ...812...65F, 2019ApJ...887..133B}). In~a recent X-ray variability study of Mrk 501 with {\it Swift}, the~ source showed the most extreme X-ray flare activity in  March--October 2014 during its 11.5 yr long monitoring period \citep{2017MNRAS.469.1655K}. During~this epoch, several short-term X-ray flares were detected with their amplitude varied by factors of 1.9-4.7 on weekly timescales. They also found a {moderate positive} correlation between the X-ray and the TeV fluxes, while no significant correlation was detected between 0.3$-$300 GeV and optical~fluxes.   

To date, out of about 3561\footnote{\url{https://www.ssdc.asi.it/bzcat/}.} blazars, TeV emissions have been detected only in 68\footnote{\url{http://tevcat.uchicago.edu/}.} blazars. Most of the TeV blazars belong to the HBL (52) subclass. It has been reported that the X-ray emissions of the TeV $\gamma-$ray emitting blazars are usually highly variable on intraday timescales (i.e.,  in less than a day time interval; (e.g., \citep{2017ApJ...841..123P,2018ApJ...859...49P}, and references therein)). High flux variability on such small timescales has been one of the most puzzling issues in the field of blazar astronomy as it requires very large energy outputs within small regions. These emitting regions often lie near to SMBH. Study of X-ray flux variability on { very short (or intraday) timescales is helpful in probing the size of the emitting regions near SMBH as well as the underlying emission mechanisms (leptonic or hadronic; (e.g., \cite{2015ApJ...811..143P})).}
The shape of X-ray spectra of TeV blazars has also been studied for quite some time as it provides information about the distribution of particles at these energies. 
The X-ray spectra of TeV blazars Mrk 421 and Mrk 501 were examined by \cite{2004A&A...413..489M,2004A&A...422..103M,2006A&A...448..861M} and they  described them in terms of the curved log-parabola model. The X-ray spectral shape of 29 TeV blazars observed with {\it Swift/XRT} were investigated by \cite{2016MNRAS.458...56W} and they found that most of them are well described by the curved log-parabola model. 
{ In our earlier work \citep{2017ApJ...841..123P}, we performed the timing analysis of five TeV blazars (including Mrk 501) using Nuclear Spectroscopic Telescope Array (NuSTAR)} data. The~number of observations of Mrk 501 studied in earlier work was only four, all taken in 2013. In~the last few years more (three) NuSTAR observations of the source were carried out. These three new observations were not studied earlier by any author. 
The motivation of this work is to examine all the {\it NuSTAR} light curves of the TeV blazar Mrk 501 for intraday flux and spectral variability, and~to investigate the nature of its 3--79 keV {X-ray spectra.} 


The outline of the manuscript is as follows. We briefly describe the {\it NuSTAR} observations and data processing in Section~\ref{sec:obs} as well as discuss the analysis techniques used for examining variability properties in Section~\ref{sec:analysis}. The~results are given in Section~\ref{sec:res}. A~summary of the work and discussion of our results are presented in Section~\ref{sec:diss}.

\section{\textbf{\textit{NuSTAR}} Observations and Data~Processing} \label{sec:obs}
{\it NuSTAR} is a space observatory consisting of two hard X-ray focusing telescopes and the two corresponding co-aligned focal plane modules, referred to as FPMA and FPMB \citep{2013ApJ...770..103H}. 
We downloaded all {\it NuSTAR} observations of the TeV blazar Mrk 501 from the HEASARC Data archive\footnote{\url{http://heasarc.gsfc.nasa.gov/docs/archive.html}.}. The~blazar Mrk 501 was observed with {\it NuSTAR} on seven occasions between 13 April, 2013 and  19 April, 2018 and the good exposure times {(those corrected for the periods of South Atlantic Anomaly (SAA) passage and Earth occultation)} ranged from 9.98 ks to 25.76 ks. In~2013, {\it NuSTAR} observed Mrk 501 four times as a part of an extensive simultaneous multiwavelength campaign \citep{2015ApJ...812...65F}. It was observed two times in 2017 in {\it NuSTAR} guest observer program cycle 2, while a single observation was made in 2018 in the extragalactic legacy survey as a part of thw {\it NuSTAR} extended mission. The~observation log of {\it NuSTAR} data for Mrk 501 is given in Table~\ref{tab:obs_log}.

\begin{table}[H]
\centering
\caption{Observation log of \textit{Nuclear Spectroscopic Telescope Array (NuSTAR)} data for TeV HBL Markarian (or Mrk)~501.}
 \label{tab:obs_log}
 \begin{tabular}{lccccc}
  \toprule
\textbf{Obs. ID}      &   \textbf{Obs. Date}  & \textbf{Start Time (UT)} &   \textbf{MJD Start}  &	\textbf{Exposure}  & \textbf{Total Elapsed}  \\
	     &  \textbf{yyyy-mm-dd}  & \textbf{hh-mm-ss}        &		     & \textbf{Time (ks)} & \textbf{Time (ks)}       \\\midrule
60002024002  &  2013-04-13  &  02:31:07       &	56395.11465  &   18.08   &   35.76 \\ 
60002024004  &  2013-05-08  &  20:01:07       &	56420.84450  &   25.76   &   55.20 \\ 
60002024006  &  2013-07-12  &  21:31:07       & 56485.90848  &   10.21   &   20.90 \\ 
60002024008  &  2013-07-13  &  20:16:07       &	56486.85136  &   ~9.98   &   20.71 \\ 
60202049002  &  2017-04-27  &  22:26:09       &	57870.94373  &   22.05   &   48.73 \\ 
60202049004  &  2017-05-24  &  22:11:09       &	57897.93135  &   23.67   &   48.30 \\ 
60466006002  &  2018-04-19  &  15:06:09       &	58227.63108  &   23.11   &   44.01 \\ 
\bottomrule
\end{tabular}\\
\end{table}

The {\it NuSTAR} data sets for the TeV blazar Mrk 501 were processed using the NuSTARDAS software within the HEASOFT\footnote{\url{http://heasarc.gsfc.nasa.gov/docs/nustar/analysis/}.} software package version 6.26.1. We first generated the calibrated and cleaned level 2 event files using the standard {\it nupipeline} script and the updated CALDB files version 20200526. For~each observation, the~source light curve and spectrum were then extracted from a circular region of radius 50$^{\prime\prime}$ centred at the source using the {\it nuproducts} script. To~extract background data for each observation we used a circular region of the similar radius on the same detector module in which the source was located but free from source~contamination. 

Since the two {\it NuSTAR} detectors, FPMA and FPMB, are co-aligned and nearly identical, we summed their background-subtracted light curves and rebinned them using a bin size of 300 s to get the final light curves. We grouped each source spectra using {\it grppha} to ensure at least 20
counts per~bin.

\section{Analysis~Techniques} \label{sec:analysis}
\unskip
\subsection{Fractional~Variance} \label{subsec:frac}
To quantify the amplitude of variability in the blazar light curves (LCs), we employed the fractional variance which is generally used for examining AGN X-ray LCs (e.g., \citep{2003MNRAS.345.1271V,2016MNRAS.458.2350W}). The~fractional variance, $F_{var}$, is~explained in detail in our previous papers (\citep{2017ApJ...841..123P,2018ApJ...859...49P}, and references therein). For~a light curve consisting of $N$ data points, the~value of $F_{var}$ is given by:
\begin{equation}
F_{var} = \sqrt{\frac{S^2 - \overline{\sigma_{err}^2}}{{\bar{x}^2}}}.
\end{equation}
The error in $F_{var}$ is calculated as: 
\begin{equation}
err(F_{var}) =  \sqrt{\left( \sqrt{\frac{1}{2N}}\frac{\overline{\sigma_{err}^2}}{\bar{x}^2 F_{var}} \right)  ^ 2+ \left(  \sqrt{\frac{\overline{\sigma_{err}^2}}{N}} \frac{1}{\bar{x}}\right) ^2 },
\end{equation}
where $S^2$ and $\bar{x}$ are the sample variance and the arithmetic mean of the LC, respectively, while $\overline{\sigma_{err}^2}$ is the mean square~error. 

\subsection{Hardness~Ratio} \label{sec:HR}
To examine spectral variations in the {\it NuSTAR} X-ray emission from the TeV blazar Mrk 501, we also extracted LCs in two energy bands: A soft energy band ranging from 3 to 10 keV and a hard energy band from 10 to 79 keV. We then calculated the hardness ratio (HR) as:
\begin{equation}
HR = \frac{H}{S},
\end{equation}
where $S$ and $H$ are the {individual simultaneous fluxes (in the units of counts/sec)} in soft and hard energy bands, respectively. The~error in HR is computed as:
\begin{equation}
\sigma_{HR} = \frac{H}{S} \sqrt {\left(\frac{\sigma_H}{H}\right)^2 + \left(\frac{\sigma_S}{S}\right)^2},
\end{equation}
where $\sigma^2_S$, and~$\sigma^2_H$ are the errors in soft and hard energy bands, respectively. {Since a hardness ratio simply compares the number of counts observed in two different energy bands, it is a model independent technique to study the spectral variability of the source.} We examined variations of the HRs with the total flux to study spectral changes with brightness in the 3--79 keV energy~range.


\section{Results}\label{sec:res}
\unskip
\subsection{X-ray Flux~Variability}\label{subsec:var}
\begin{figure}[H]
\centering
\includegraphics[width=15.5cm]{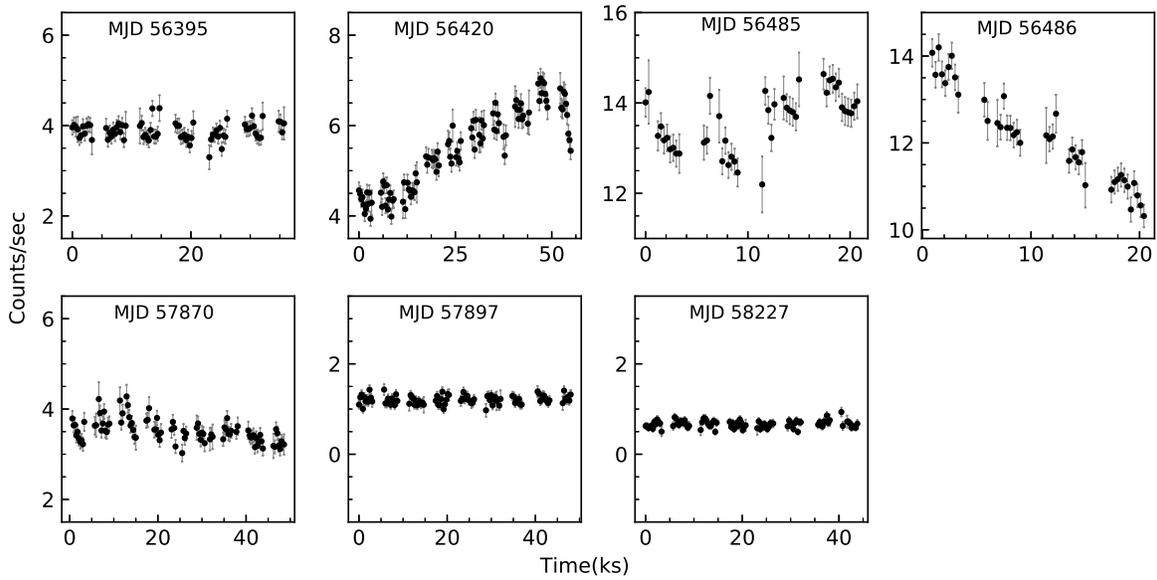}
\caption{\textit{NuSTAR} light curves of the TeV blazar Mrk 501. The MJD of observation is given in each~plot.} 
\label{fig:lc} 
\end{figure}

We plotted all the {\it NuSTAR} light curves of the TeV blazar Mrk 501 in Figure~\ref{fig:lc}. Clear intraday 
variations were seen on a couple of nights of observations. We examined all the light curves of Mrk 501 for intraday flux variations using fractional variability amplitude F$_{\rm var}$, discussed in Section~\ref{subsec:frac}. The~results of our intraday variability (IDV) analyses are given in Table~\ref{tab:var_par}, where dashes ``--'' indicate that the  sample variances $<$ mean square errors, so no real value of $F_{var}$ can be computed for the LC. The~source showed significant flux variations on four out of seven nights. The~maximum flux variation was detected on MJD 
56420 (Obs. ID: 60002024004) with fractional variability amplitude of 15.53\%, while the minimum flux variation was observed on MJD 56485 (Obs. ID: 60002024006) with a F$_{\rm var}$ value of 3.90\%. The~values of F$_{\rm var}$ were 8.45\%, and~5.17\% on MJD 56486 and~MJD 57870, respectively. No significant IDV variation was detected on MJD 56395, MJD 57897, and~MJD 58227. We also examined the soft (3--10 keV) energy band LCs and the hard (10--79 keV) energy band LCs for intraday variability. As~can be seen from the Table~\ref{tab:var_par}, the~amplitude of variability is larger in the higher (10--79 keV) energy band than the lower (3--10 keV) energy band during all the observations with significant IDV detection in 3--79 keV X-ray~LCs.

\begin{table}[H]
\centering
\caption{Hard X-ray variability properties of Mrk~501.}
\label{tab:var_par}
\begin{tabular}{lccccccc}
  \toprule
\textbf{Obs. ID}         &     \textbf{MJD}	&            \multicolumn{3}{c}{ \boldmath$F_{var}(\%)$}                      \\
	        &		&  \textbf{Soft (3--10 keV)}      & \textbf{Hard (10--79 keV)}      & \textbf{Total (3--79 keV) }	\\ \midrule
60002024002     &   56,395	& $         -         $ & $         -         $ & $         -         $   \\ 
60002024004     &   56,420	& $  14.11 \pm   0.45 $ & $  21.33 \pm   0.92 $ & $  15.53 \pm   0.40 $   \\ 
60002024006     &   56,485	& $   3.29 \pm   0.54 $ & $   5.45 \pm   1.06 $ & $   3.90 \pm   0.44 $   \\ 
60002024008     &   56,486	& $   7.75 \pm   0.50 $ & $  10.63 \pm   0.98 $ & $   8.45 \pm   0.44 $   \\ 
60202049002     &   57,870	& $   5.02 \pm   0.75 $ & $   7.08 \pm   1.79 $ & $   5.17 \pm   0.64 $   \\ 
60202049004     &   57,897	& $         -         $ & $         -         $ & $         -         $   \\ 
60466006002     &   58,227	& $   5.10 \pm   2.45 $ & $         -         $ & $   4.14 \pm   2.59 $   \\ 

\bottomrule
\end{tabular}
\end{table}

\subsection{X-ray Spectral~Variability}\label{subsec:spec_var}
\begin{figure}[H]
\centering
\includegraphics[width=15.5cm]{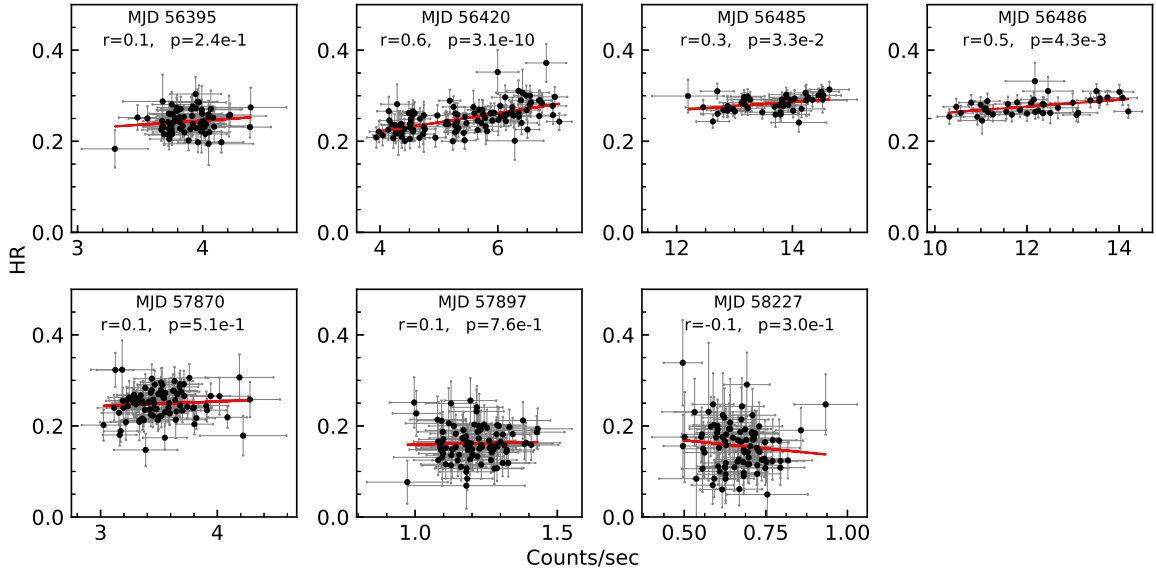}
\caption{\label{fig:hr}The variation of hardness ratio (HR) with respect to 3--79 keV flux (in the units of count rates). The~{ MJD of observation}, correlation coefficient (r), and the null hypothesis probability (p) are given in each~plot.} 
\end{figure}

We investigated the spectral variability of 3--79 keV X-ray emission from the TeV blazar Mrk 501 using a model-independent hardness-ratio analysis. We plotted the HR against total (3--79 keV) flux in Figure~\ref{fig:hr}. To~search for any systematic variation in HR with flux, we fitted a first-order polynomial in each plot of Figure~\ref{fig:hr}. The~values of correlation coefficient (r) and the null hypothesis probability (p) are mentioned in each plot. { Assuming Gaussian fluctuations and white noise variations, strong }positive correlations (r $\ge$ 0.5 and p < 0.01) between HR and total count rates were found on MJD 56420 (Obs. ID: 60002024004; \mbox{r = 0.6}, {p} = 3.1 $\times$ 10$^{-10}$) and MJD 56486 (Obs. ID: 60002024008; \mbox{r = 0.5}, \mbox{{p} = 4.3 $\times$ 10$^{-3}$}) indicating that the spectra become harder with increasing flux during these observations. No~statistically significant correlations were detected in the remaining five~observations.

\subsection{{\it~NuSTAR} Spectra}\label{subsec:spectra}
For each observation, we simultaneously fitted the spectra from the two {\it NuSTAR} detectors, FPMA and FPMB, in~the XSPEC\footnote{\url{https://heasarc.gsfc.nasa.gov/docs/xanadu/xspec/XspecManual.pdf}.} version 12.10.1f using $\chi^2$ statistics. To~account for cross-calibration uncertainties between these two detectors we have included a multiplicative constant factor in each spectral model whose value is kept fixed to 1 for FPMA and free to vary for FPMB. The~values of the constant factor ranged from 0.97 to 1.05 which is within the expected values as suggested by~\cite{2015ApJS..220....8M}.

The spectra were modelled with both a simple power-law (PL) and a log-parabolic (LP) model. The~power-law model is defined as:
\begin{equation}
F(E) = K E^{-\Gamma}
\end{equation}
where $\Gamma$ and $K$ are the photon-index and the normalization, respectively, while $F(E)$ is the flux at energy $E$. The~log-parabolic model is given by \citep{2004A&A...413..489M}:
\begin{equation}
F(E) = K (E/E_{pivot})^{-(\alpha+\beta \log(E/E_{pivot}))}
\end{equation}
where $\alpha$ is the photon-index at fixed energy $E_{pivot}= 10$ keV, $\beta$ is the spectral curvature, and~K is the~normalization.

We multiplied each spectral model by a \textit{phabs} component with fixed hydrogen column density, taken from~\cite{2005A&A...440..775K}, to~account for the Galactic~absorption. 

To choose the best-fitted spectral model between PL and LP we performed the {\it F}-test\footnote{F-test is available in the XSPEC.} using  values of $\chi^2$ and the dof\footnote{Degree of freedom.} for both  models.  
The results of spectral fitting and the {\it F}-test are listed in Table~\ref{tab:spectra}. The~LP model provides a better fit than the simple PL model
if the value of {\it F}-statistic $>$ 1 and the corresponding null hypothesis probability, {\it p} $<$ 0.01. 
We found that for five out of seven {\it NuSTAR} observations of Mrk 501 the curved LP model provides a better fit over the PL model. For~the last two observations, the~{\it NuSTAR} spectra of Mrk 501 are well described by the PL model as can be seen from the high values ($p > 0.01$) of null hypothesis probability. The~model-fitted spectra and the data-to-model ratios for each observation are plotted in Figure~\ref{fig:spectra}.

\begin{table}[H]
\centering
\caption{Model fits to the NuSTAR spectra of Mrk~501.}
\label{tab:spectra}
\scalebox{.8}[0.8]{\begin{tabular}{lcccccccc}
  \toprule
\textbf{Obs. ID}      & \multicolumn{2}{c}{\textbf{Power Law}}                  & \multicolumn{3}{c}{ \textbf{Log-Parabola (\boldmath$E_{pivot}=10$ keV)} }            & \textbf{Flux}\boldmath$_{\text{3--79 keV}}\,^{(1)}$      & \textbf{F-test}     & \textbf{\emph{p}-Value}   \\
\cmidrule[0.03cm](r){2-3}\cmidrule[0.03cm](r){4-6}	
             &      \boldmath$\Gamma $      &  \boldmath$ \chi^2/dof $($\chi^2_r$) &  \boldmath$\alpha$          &  \boldmath$\beta $          &  \boldmath$ \chi^2/dof $($\chi^2_r$) &  &&\\
               
  \midrule
60002024002  & $ 2.28 \pm 0.02 $ & ~683.48/651 (1.04)           & $ 2.32 \pm 0.02 $ & $ 0.21 \pm 0.06 $ & 649.73/650 (0.99)   	     & $ ~8.80 \pm 0.20 $ &$ ~33.76 $ &  9.75 $\times$ 10 $^{-9}$  \\

60002024004  & $ 2.24 \pm 0.01 $ & ~937.49/834 (1.12) 	        & $ 2.26 \pm 0.01 $ & $ 0.15 \pm 0.04 $ &  899.86/833 (1.08)  	     & $ 13.13 \pm 0.21 $ &$ ~34.83 $ & 5.22 $\times$  10$^{-9}$  \\

60002024006  & $ 2.10 \pm 0.01 $ & 1002.01/865 (1.15)      	& $ 2.12 \pm 0.01 $ & $ 0.21 \pm 0.04 $ &  917.77/864 (1.06)         & $ 35.31 \pm 0.57 $ &$ ~79.30 $ & 3.09 $\times$  10$^{-18}$  \\

60002024008  & $ 2.12 \pm 0.01 $ & ~903.40/804 (1.12)		& $ 2.16 \pm 0.01 $ & $ 0.28 \pm 0.04 $ &  767.64/803 (0.95)         & $ 30.53 \pm 0.51 $ &$ 142.01 $ & 2.90 $\times$  10$^{-30}$  \\

60202049002  & $ 2.21 \pm 0.02 $ & ~738.43/670 (1.10)		& $ 2.23 \pm 0.02 $ & $ 0.16 \pm 0.06 $ &  717.37/669 (1.07)         & $ ~8.74 \pm 0.20 $ &$ ~19.64 $ &  1.092 $\times$  10$^{-5}$ \\

60202049004  & $ 2.70 \pm 0.03 $ & ~390.07/406 (0.96)		& $ 2.76 \pm 0.05 $ & $ 0.19 \pm 0.16 $ &  384.27/405 (0.94)         & $ ~2.37 \pm 0.05 $ &$ ~~6.11 $ & $ 0.02 $ \\

60466006002  & $ 2.75 \pm 0.04 $ & ~268.94/302 (0.89)		& $ 2.82 \pm 0.08 $ & $ 0.21 \pm 0.20 $ &  265.76/301 (0.88)         & $ ~1.25 \pm 0.04 $ &$ ~~3.60 $ & $ 0.06 $ \\
	
\bottomrule
\end{tabular}}\\
\begin{tabular}{lcccccccc}
\multicolumn{1}{c}{\footnotesize $^{(1)}$ 3--79 keV unabsorbed flux (in units of $ 10^{-11}$ erg cm$^{-2}$ s$^{-1}$) for the best fitted model. }
\end{tabular}
\end{table}

\begin{figure}[H]
\centering
\begin{tabular}{ccc}
    \includegraphics[width=7cm,height=7cm]{lp_60002024002}&
    \includegraphics[width=7cm,height=7cm]{lp_60002024004}\\[1\tabcolsep]
    \includegraphics[width=7cm,height=7cm]{lp_60002024006}&
    \includegraphics[width=7cm,height=7cm]{lp_60002024008}\\[1\tabcolsep]
    \multicolumn{2}{c}{(\textbf{a})}\\
    \end{tabular}
\caption{\textit{Cont}.}
\end{figure}

\begin{figure}[H]
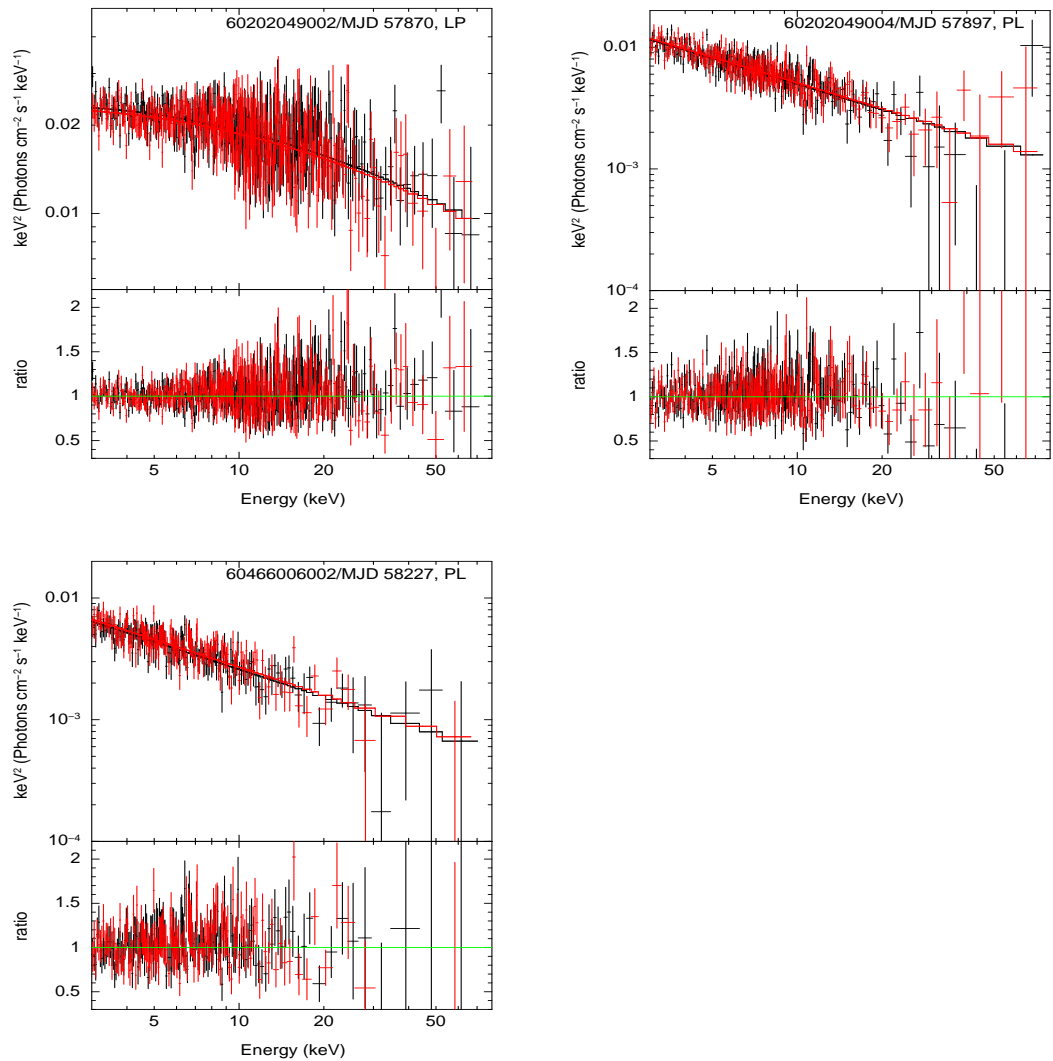
\ContinuedFloat
\centering
\begin{tabular}{ccc}
    \includegraphics[width=7cm,height=7cm]{lp_60202049002}&
    \includegraphics[width=7cm,height=7cm]{pl_60202049004}\\[1\tabcolsep]
    \includegraphics[width=7cm,height=7cm]{pl_60466006002}&\\
        \multicolumn{2}{c}{(\textbf{b})}\\
    \end{tabular}
\caption{\label{fig:spectra}(\textbf{a},\textbf{b}) \textit{NuSTAR} spectra of TeV HBL Mrk 501 with the best fitting model in the upper panel and the ratio (data/model) in the bottom panel of each plot. The~observation ID, {MJD of observation}, and the best fitting model are given in each~plot.}
\end{figure}

\section{Summary and~Discussion}\label{sec:diss}
In this paper, we analysed seven archival {\it NuSTAR} observations of the TeV blazar Mrk 501 performed between  13 April 2013 and 19 April 2018. In~particular, we studied hard (3--79 keV) X-ray flux and spectral variability properties of Mrk 501, and~investigated the shape of its X-ray spectra in 3--79 keV energy range. We examined all {\it NuSTAR} light curves of Mrk 501 for intraday variability using fractional variability analysis. We found significant IDV variations in four out of seven X-ray LCs with the fractional variability amplitudes ranging from 3.90\% to 15.53\%. No significant variations were detected in the rest three observations. {We noticed that during the last two observations (MJD 57897, and~MJD 58227) the source was in somewhat lower states, while on MJD 56395 the intrinsic variations in the light curve are probably smaller than the measurement errors.} 

TeV blazars are known to exhibit large amplitude flux variations at all frequencies on different timescales (e.g., \citep{2008MNRAS.386L..28G, 2015ApJ...811..143P, 2017ApJ...841..123P,2018ApJ...859...49P,2019ApJ...871..192P,2020ApJ...890...72P}, and references therein). Flux variations in blazar LCs are generally understood to originate from the relativistic jets \citep{2014ApJ...780...87M}. At~high frequencies (X-rays to $\gamma-$rays) large amplitude variations are often observed on very short timescales indicating compact emitting regions (e.g., \citep{2007ApJ...664L..71A, 2007ApJ...669..862A}). Such rapid variations are {often} explained by the interaction of relativistic turbulent plasma with shock within the jets, which accelerates the particles to high energies \citep{2014ApJ...780...87M}. The~shortest hard X-ray (3--79 keV) variability with doubling time of $\sim 14$ minutes was reported from the TeV blazar Mrk 421 by~\cite{2015ApJ...811..143P} during its 2013 April flare. They suggested that magnetic reconnections taking place in fast-moving emitting regions within the jets could be responsible for such a short timescale variability (``jets-in-a-jet'' model; \citep{2009MNRAS.395L..29G}). 
{Long term flux variations in the X-ray light curve of Mrk 501 during the entire {\it NuSTAR} monitoring period can be} explained by the shock-in-jet model which involves the acceleration of particles to high energies by an internal shock within the jet followed by subsequent cooling by emitting radiations \citep{1985ApJ...298..114M}. Other possible explanations for the longer timescale variations include a change in the viewing angle, and~hence in the Doppler factor of the emitting region within the jet (e.g., \citep{1992A&A...255...59C, 1999A&A...347...30V}), and/or a change in the magnetic field \citep{2010ApJ...725.2344B}. 

We examined the hard (3--79 keV) X-ray spectral variability of the TeV HBL Mrk 501 using HR analysis. We found that for observations {on MJD 56420 and~MJD 56486}, the HR increases with increasing flux, that is, the~{\it NuSTAR} spectra of Mrk 501 became harder with the increasing count rates. Such a ``harder-when-brighter'' trend has often been noticed in X-ray observations of the HBL-type blazars (e.g., \citep{2002ApJ...572..762Z, 2018MNRAS.480.4873A, 2017ApJ...841..123P}, and references therein). For~the other five observations, no strong correlation were seen between HR and count rates. {Moreover, as~can be seen from Figure~\ref{fig:hr}, the~data have uncertainties 
which were not taken into account during the correlation analysis and thus, the~correlation between the HR and the flux may be somewhat lower than those reported here.}

We also studied the shape of {\it NuSTAR} spectra of TeV blazar Mrk 501 using the simple power-law model and the log-parabola model. We found that  five out of seven {\it NuSTAR} spectra are well described by the curved LP models with local photon indices $\alpha$ lying in the range of 2.12--2.32 and the spectral curvature $\beta \sim$ 0.15--0.28. For~the last two observations (Obs. IDs: 60202049004 and 60466006002), when the X-ray fluxes are relatively low the simple PL model provides an equivalently good fit with photon indices $\Gamma \sim$ 2.70 and 2.75, respectively. During~the {\it NuSTAR} monitoring period of Mrk 501, the maximum 3--79 keV unabsorbed flux recorded was $35.31 \times 10^{-11}$ erg cm$^{-2}$ s$^{-1}$ on MJD 56485, while the minimum flux observed was $1.25 \times 10^{-11}$ erg cm$^{-2}$ s$^{-1}$ on MJD~58227. 

The study of X-ray spectral shape of TeV blazars is helpful in understanding the particle acceleration mechanism and the distribution of emitting particles. The~curved LP model was used for the first time to describe the spectral shape of synchrotron emission from the BL Lac type blazars by~\cite{1986ApJ...308...78L}. Later,~\cite{2004A&A...413..489M,2004A&A...422..103M,2006A&A...448..861M} successfully described the X-ray spectra of TeV HBLs Mrk 421 and Mrk 501 using the LP model and suggested that the curved X-ray spectra of TeV blazars can be understood in terms of statistical energy-dependent particle acceleration. The~observed curvature of X-ray spectra of TeV HBLs could be either convex or concave. The convex curvature of the X-ray spectra, as~found in this work, is likely to be caused by a single-accelerated particle distribution (e.g., \citep{2004A&A...413..489M}), while the concave X-ray spectra observed in some studies (e.g., \citep{2008ApJ...682..789Z}), maybe a consequence of the spectral upturn at the interaction of the high-energy tail of the synchrotron emission and the low-energy part of the IC~emission. 



\acknowledgments{{This research has made use of data obtained with NuSTAR, the~first focusing hard X-ray mission managed by the Jet Propulsion Laboratory (JPL), and~funded by the National Aeronautics and Space Administration (NASA). This research has also made use of the NuSTAR Data Analysis Software (NuSTARDAS) jointly developed by the ASI Science Data Center (ASDC, Italy) and the California Institute of Technology (Caltech, USA).}}



\reftitle{References}

\end{document}